\newcommand{\NPA}[3]{Nucl.\ Phys.\ A\ {\bf #1},\ #2 (#3)}
\newcommand{\NPB}[3]{Nucl.\ Phys.\ B\ {\bf #1},\ #2 (#3)}

\newcommand{\PLB}[3]{Phys.\ Lett.\ B\ {\bf #1},\ #2 (#3)}
\newcommand{\PR}[3]{Phys.\ Rep.\ {\bf #1},\ #2 (#3)}
\newcommand{\PRL}[3]{Phys.\ Rev.\ Lett.\ {\bf #1},\ #2 (#3)}

\newcommand{\PRC}[3]{Phys.\ Rev.\ C\ {\bf #1},\ #2 (#3)}
\newcommand{\PRD}[3]{Phys.\ Rev.\ D\ {\bf #1},\ #2 (#3)}

\newcommand{\EPJC}[3]{Eur.\ Phys.\ J.\ C\ {\bf #1},\ #2 (#3)}
\newcommand{\EPJA}[3]{Eur.\ Phys.\ J.\ A\ {\bf #1},\ #2 (#3)}

\newcommand\e{\epsilon}

\newcommand\f{\phi}

\newcommand{\diracslash}[1]{#1\llap{/\kern2pt}}

\newcommand{\be}{\begin{equation}}
\newcommand{\ee}{\end{equation}}
\newcommand{\bea}{\begin{eqnarray}}
\newcommand{\eea}{\end{eqnarray}}
\newcommand{\ba}[1]{\begin{array}{#1}}
\newcommand{\ea}{\end{array}}

\documentclass[prd,aps,floats,nofootinbib,tightenlines,showpacs]{revtex4-1} 
\usepackage{graphicx}
\usepackage{epsfig,graphicx,pstricks}
\usepackage{wrapfig}
\usepackage{psfrag}
\usepackage{color}
\usepackage{amsmath}
\usepackage{amsfonts}
\usepackage{amssymb}
\usepackage{textcomp}
\usepackage{axodraw}
\usepackage{multirow}
\usepackage{float}

\begin{document}

\title { Transport coefficients in the Polyakov quark meson coupling model:
 A quasi particle approach}
\author{Aman Abhishek }
\email{aman@prl.res.in}
\affiliation{Theory Division, Physical Research Laboratory,
Navrangpura, Ahmedabad 380 009, India}
\author{Sabyasachi Ghosh}
\affiliation { Department of Physics, University of Calcutta, 92, A.P.C. Road, Kolkata 700009, India}
\email{sabyaphy@gmail.com}
\author{Hiranmaya Mishra}
\email{ Speaker, hm@prl.res.in}
\affiliation{Theory Division, Physical Research Laboratory,
Navrangpura, Ahmedabad 380 009, India}
%
%
\def\be{\begin{equation}}
\def\ee{\end{equation}}
\def\bearr{\begin{eqnarray}}
\def\eearr{\end{eqnarray}}
\def\zbf#1{{\bf {#1}}}
\def\bfm#1{\mbox{\boldmath $#1$}}
\def\hf{\frac{1}{2}}
\def\sl{\hspace{-0.15cm}/}
\def\omit#1{_{\!\rlap{$\scriptscriptstyle \backslash$}
{\scriptscriptstyle #1}}}
\def\vec#1{\mathchoice
        {\mbox{\boldmath $#1$}}
        {\mbox{\boldmath $#1$}}
        {\mbox{\boldmath $\scriptstyle #1$}}
        {\mbox{\boldmath $\scriptscriptstyle #1$}}
}

\begin{abstract}
We compute the transport coefficients, namely, the coefficients of shear and bulk viscosities as well as thermal
conductivity  for hot and dense matter. The calculations are performed 
within the Polyakov loop extended quark meson model.
The estimation of the transport coefficients is made using 
the Boltzmann kinetic equation within the relaxation time approximation. 
The energy dependent relaxation time is estimated from meson meson 
scattering, quark meson scattering and quark quark scattering within the model.
In our calculations, the shear viscosity to entropy ratio and
the coefficient of thermal conductivity show a minimum at the critical
temperature, while the ratio of bulk viscosity to entropy density exhibits a
peak at this transition point.
 The effect of confinement modelled through a 
Polyakov loop potential plays an important role in the estimation of these dissipative coefficients
both below and above the critical temperature.

\vspace{2in}
\begin{center}
(To appear in Proceedings of Science, Proceedings of 
"Critical Point and Onset of Deconfinement - CPOD2017
7-11 August, 2017
The Wang Center, Stony Brook University, Stony Brook, NY)
\end{center}

%
\end{abstract}
%

\maketitle

\section{Introduction}

 Transport coefficients of matter under extreme conditions of temperature, density or
external fields are interesting  and important  for several reasons.
In the context of relativistic 
heavy ion collisions, these properties enter as dissipative coefficients in the 
hydrodynamic evolution of the quark gluon plasma. They are also important for the
cooling of neutron stars. 
The cooling
of neutron stars at short time scales constrains the thermal conductivity 
\cite{chamel} while the cooling through
neutrino emission on a much larger time scales constrains the phase of the matter in the interior of the compact star 
\cite{yakovlev}. 
 Apart from these,
the temperature and chemical potential dependence of the transport coefficients may actually reveal the location of phase transitions
\cite{kapustabk}. 

Transport coefficients for QCD matter in principle can be calculated using Kubo formulation \cite{kubo}. However, QCD is 
strongly interacting for both at energies accessible in heavy ion collision experiments as well as for the
densities expected to be there in the core of the neutron stars making the perturbative estimations unreliable.
Calculations using lattice QCD simulations at finite chemical potential is also challenging and is limited only
to the equilibrium thermodynamic properties at small chemical potentials. This has motivated to estimate the
transport coefficient in various effective models of strong interaction physics. These include
chiral perturbation theory \cite{dobadoch}, quasi-particle models \cite{quasip}, linear sigma model \cite{purnendu} and the Nambu-
Jona-Lasinio model \cite{sasakinjl,deb}.  The general temperature dependence of the viscosity coefficients turns out to be
similar with the ratio of shear viscosity to entropy density ($\eta/s$)  exhibiting a minimum at the transition temperature.
 The numerical value of $\eta$  at the minimum however differ by order of magnitude.
Similarly, the bulk viscosity shows a maximum near the critical temperature. 
The numerical values
of these coefficients however, vary over  a large range of values e.g. 
 $\zeta$ varies from $10^{-5}$ GeV$^3$ \cite{souravsuk} to $10^{-2}$ GeV$^3$ around the critical temperature
\cite{sasakinjl}.

The other transport coefficient that is important at finite baryon density is the coefficient of
thermal conductivity $\lambda$ \cite{denicolhydro,denicolpre,denicolheat}. 
This coefficient has been evaluated in various
effective models like Nambu Jona Lasinio model using Green-Kubo approach \cite{fukutome},
relaxation time approximation \cite{deb}
and the  instanton liquid model \cite{nam}. The results, however,
 vary over a  wide range of  values, with 
$\lambda=0.008$ GeV$^{-2}$ as in Ref. \cite{nicola} 
to $\lambda \sim 10$ GeV$^{-2}$ as in Ref. \cite{marty} for a 
range of temperatures (0.12 GeV $<$T$<$ 0.17 GeV),  which  has 
been nicely tabulated in 
Ref. \cite{sabyath}. 

We shall here attempt to estimate these transport coefficients within
an effective model of strong interaction, the
Polyakov loop extended quark meson (PQM) model that incorporates the aspects of chiral symmetry breaking
in strong interaction and takes care of confinement deconfinement transition partially while explicitly keeping pionic degrees of freedom at low temperature.

 The transport coefficients are evaluated within the
relaxation time approximation of Boltzmann equation which is a reasonable approximation for quasi particles \cite{quasip,guru}.
 The relaxation time 
is calculated from the scattering of the particles that constitute the
dynamical degrees of freedom of the model - namely the 
meson scattering, as is considered in Ref.\cite{purnendu} with medium
dependent meson masses; quark scattering through meson exchanges
similar to as considered in Ref.s \cite{sasakinjl,deb,marty} with medium 
dependent quark and meson masses and quark meson scattering.
As we shall see in the following, each of these processes bring out distinct
features for the transport coefficients.


\section{Thermodynamics of PQM model} 
 The thermodynamic potential in PQM model is given by\cite{bjschaefer,bielich,buballa,ranjita}
\be
\Omega(T,\mu)=\Omega_{\bar q q}+U_\chi+U_P(\phi,\bar\phi)
\label{thpot}
\ee
The fermionic (quark) part of the thermodynamic potential is given as
\bearr
\Omega_{\bar q q}&=&-2N_fT\int \frac{d^3 p}{(2 \pi)^3} 
\bigg[\ln\left(1+3(\phi+\bar\phi e^{-\beta\omega_-})e^{-\beta\omega_-}+e^{-3\beta\omega_-}\right)\nonumber\\
&+&\ln\left(1+3(\bar\phi+\phi e^{-\beta\omega_+})e^{-\beta\omega_+}+e^{-3\beta\omega_+}\right)\bigg]
\eearr
modulo a divergent vacuum part. In the above, $\omega_\mp=E_p\mp\mu$, with the single particle quark/anti-quark energy $E_p=\sqrt{\zbf p^2+M^2}$.
The constituent quark/anti-quark mass is defined to be
\be
M^2=g_\sigma^2(\sigma^2+\zbf \pi^2).
\ee
 In Eq.(\ref{thpot}), 
potential $U_\chi(\sigma,\zbf \pi)$ is the mesonic potential that essentially describes 
the chiral symmetry breaking pattern in strong interaction and is given by
\be
U_\chi(\sigma,\bfm \pi)=\frac{\lambda}{4}(\sigma^2+\bfm \pi^2-v^2)-c\sigma
\label{uchi}
\ee
while, the last term in Eq.(\ref{thpot}) is the Polyakov loop potential that essentially describes the confinement
deconfinement transition. 
polynomial parametrization \cite{bjschaefer}
\be
U_P(\phi,\bar\phi)=T^4\left[-\frac{b_2(T)}{2}\bar\phi\phi-\frac{b_3}{2}(\phi^3+\bar\phi^3)+\frac{b_4}{4}(\bar\phi\phi)^2\right],
\label{uphi}
\ee
with the temperature dependent coefficient $b_2$ given as
\be
b_2(T)=a_0+a_1(\frac{T_0}{T})+a_2(\frac{T_0}{T})^2+a_3(\frac{T_0}{T})^3
\ee
The numerical values of the parameters are
$
a_0=6.75, a_1=-1.95, a_2=2.625, a_3=-7.44,
b_3=0.75, b_4=7.5$
The parameter $T_0$ corresponds to the transition temperature of Yang-Mills theory. However, for the full
dynamical QCD, there is a flavor dependence on $T_0(N_f)$. For two flavors we take it to be $T_0(2)=192$ MeV as in
Ref.\cite{bjschaefer}. The parameters of potential $U_\chi$,are so chosen that the chiral symmetry is broken 
spontaneously in vacuum  with $\langle\sigma\rangle=f_\pi$, and $\langle\bfm \pi\rangle=0$ with $f_\pi=93$MeV  is the pion decay constant. The coefficient of symmetry breaking term is fixed from PCAC so that $c=f_\pi^2m_\pi^2$; $v^2=f_\pi^2-m_\pi^2/\lambda$, with
$\lambda$ determined from mass of the $\sigma$ meson  leading to $\lambda=19.7$ and $g_\sigma=3.3$ so that the constituent
quark mass in the vacuum is about 300 MeV\cite{rischkeqm}.
The mean fields are obtained by minimizing $\Omega$ with respect to $\sigma$, $\phi$, $\bar\phi$, and $\pi$. For example, extremising
 the effective potential with respect to $\sigma$ field leads to
\be
\lambda(\sigma^2+\zbf\pi^2-v^2)-c+g_\sigma\rho_s=0
\label{gapsigma}
\ee
where, the scalar density $\rho_s=-\langle\bar\psi\psi\rangle$ is given by
\be
\rho_s=6N_fg_\sigma\sigma\int\frac{d\zbf p}{(2\pi)^3}\frac{1}{E_P}\left [f_-(\zbf p)+f_+(\zbf p)\right].
\label{rhos}
\ee
In the above, $f_\mp(\zbf p)$ are the distribution functions for the quarks and anti-quarks, with $\omega_\mp=E(\zbf p)\mp\mu$,
 given as
\be
f_-(\zbf p)=
\frac{\phi e^{-\beta\omega_-}+2\bar\phi e^{-2\beta\omega_-} 
+ e^{-3\beta\omega_-}}
{1+3\phi e^{-\beta\omega_-}+3\bar\phi e^{-2\beta\omega_-} + e^{-3\beta\omega_-}},
\label{fq}
\ee
and,
\be
f_{+}(\zbf p)=\frac{\bar\phi e^{-\beta\omega_+}+2\phi e^{-2\beta\omega_+} 
+ e^{-3\beta\omega_+}}{1+3\bar\phi e^{-\beta\omega_+}+3\phi e^{-2\beta\omega_+} + e^{-3\beta\omega_+}}.
\label{fqbar}
\ee
It can be shown that  for vanishing chemical potential, $\phi=\bar\phi$ and the distribution functions become
\be
f_\phi(\zbf p)=
\frac{\phi e^{-\beta E_-}+2\phi e^{-2\beta E} 
+ e^{-3\beta E}}
{1+3\phi e^{-\beta E}+3\phi e^{-2\beta E} + e^{-3\beta E}},
\label{fphi}
\ee
where, $E(\zbf p)$ is the single particle energy for the quarks.

The meson masses for $\sigma$ and $\pi$ are  determined by the curvature of $\Omega$ at the global minimum
\be
M_\sigma^2=\frac{\partial^2\Omega}{\partial\sigma^2}|_{\sigma=\sigma_0,\pi=0},\quad\quad M_{\pi_i}^2=\frac{\partial^2\Omega}{\partial\pi_i^2|_{\sigma=\sigma_0,\pi=0}}.
\ee

The energy density $\epsilon=\Omega-T\partial\Omega/\partial T+\mu\rho_q$ is given by
\be
\epsilon=\frac{6}{\pi^2}\int p^2dp E(\zbf p)\left(f_-(\zbf p)+f_+(\zbf p)\right)+U_\chi-3 U_P(\phi,\bar\phi)
+\frac{T^5}{2}\frac{db_2(T)}{dT}\bar\phi \phi
\label{energy}
\ee
In Fig.1(a), we have plotted the constituent quark mass, and the meson masses in the model as a function of temperature for vanishing baryon density. In the chirally broken phase, m$_\pi$, being the mass
of an approximate Goldstone mode is protected and varies weakly with temperature. On the other hand,
the mass of $\sigma$ , $M_\sigma$,
which is approximately twice the constituent quark mass,$M$ drops significantly near the
crossover temperature. At high temperature, being chiral partners, the masses of $\sigma$ and $\pi$ mesons become degenerate and increase linearly with temperature.
 In Fig. 1b, we have plotted the order parameters $\sigma$ and $\phi$ as a function of temperature for vanishing quark chemical potential.
 We also note that for $\mu=0$, the order parameters $\phi$ and $\bar\phi$ are the same. 
Because of the approximate chiral symmetry, the chiral order parameter  
decreases  with temperatures to  small values but never vanishes. The Polyakov loop parameter on the other hand 
grows from $\phi=0$ at zero temperature to about $\phi=1$ at high temperatures. We might mention here that at very high temperature
exceeds unity, the value in the infinite quark mass limit.

\begin{figure}[t!]
\vspace{-0.4cm}
\begin{center}
\begin{tabular}{c c}
\includegraphics[width=9cm,height=9cm]{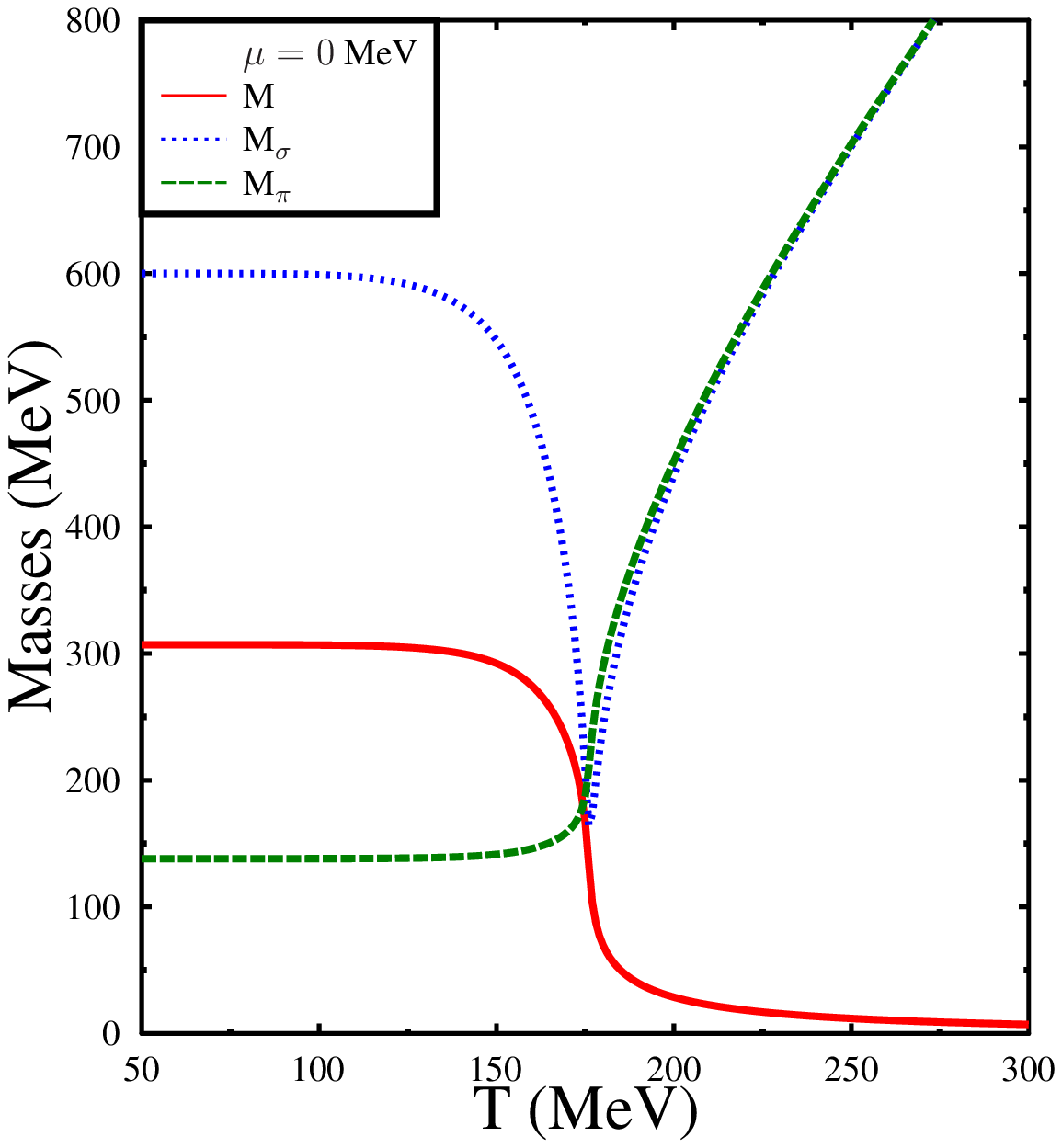}&
\includegraphics[width=9cm,height=9cm]{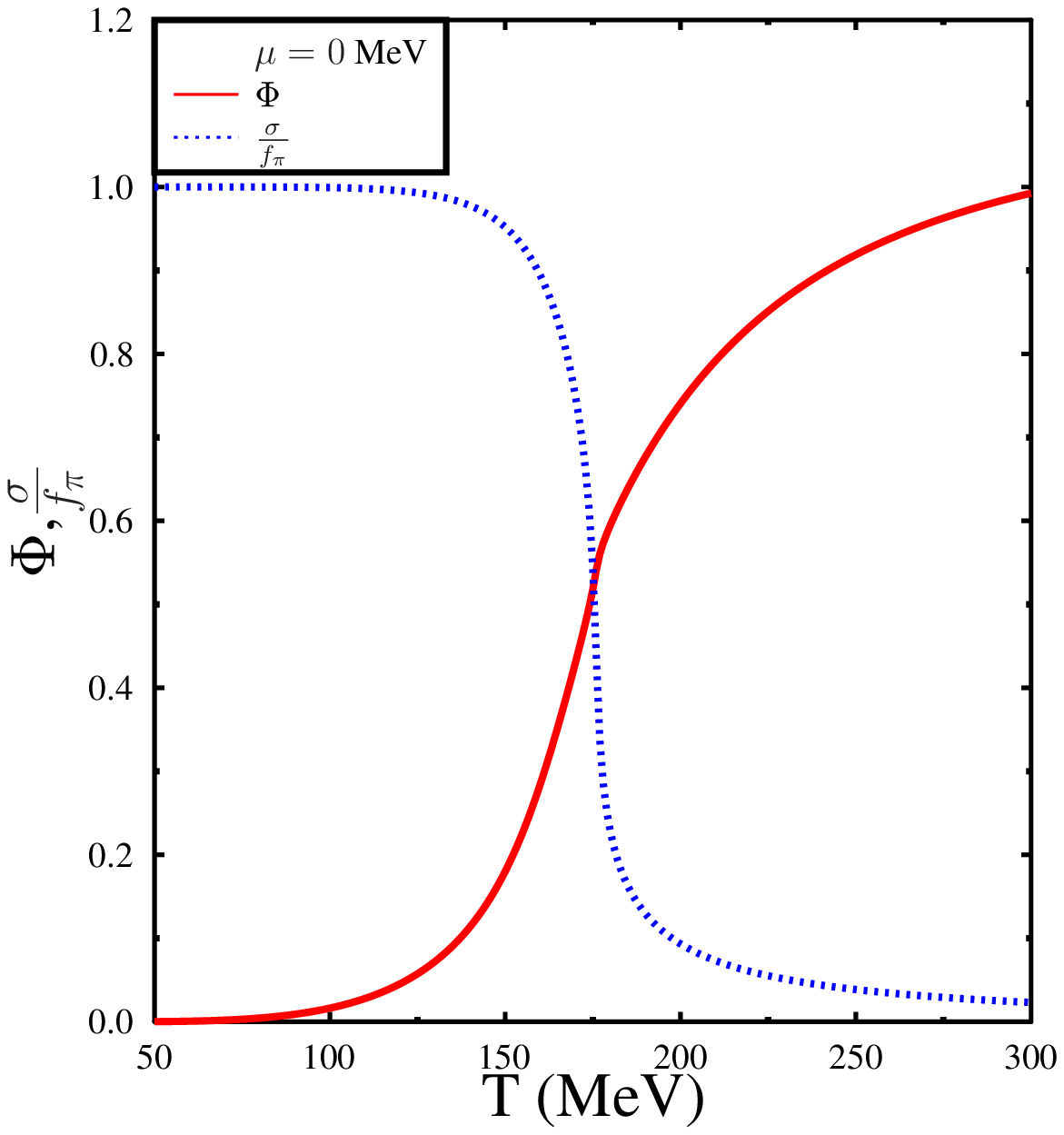}\\
Fig. 1-a & Fig. 1-b
\end{tabular}
\end{center}
\caption{ (Fig 1 a) Temperature dependence of the masses of constituent quarks ($M$), and pions ($M_\pi$) and sigma mesons ($M_\sigma$)
 and (Fig1-b)
 the order parameters $\sigma$ and $\phi$ as a function of temperature for  $\mu=0$ MeV .}
\label{fig1}
\end{figure}
\begin{figure}[t!]
\vspace{-0.4cm}
\begin{center}
\begin{tabular}{c c}
\includegraphics[width=9cm,height=9cm]{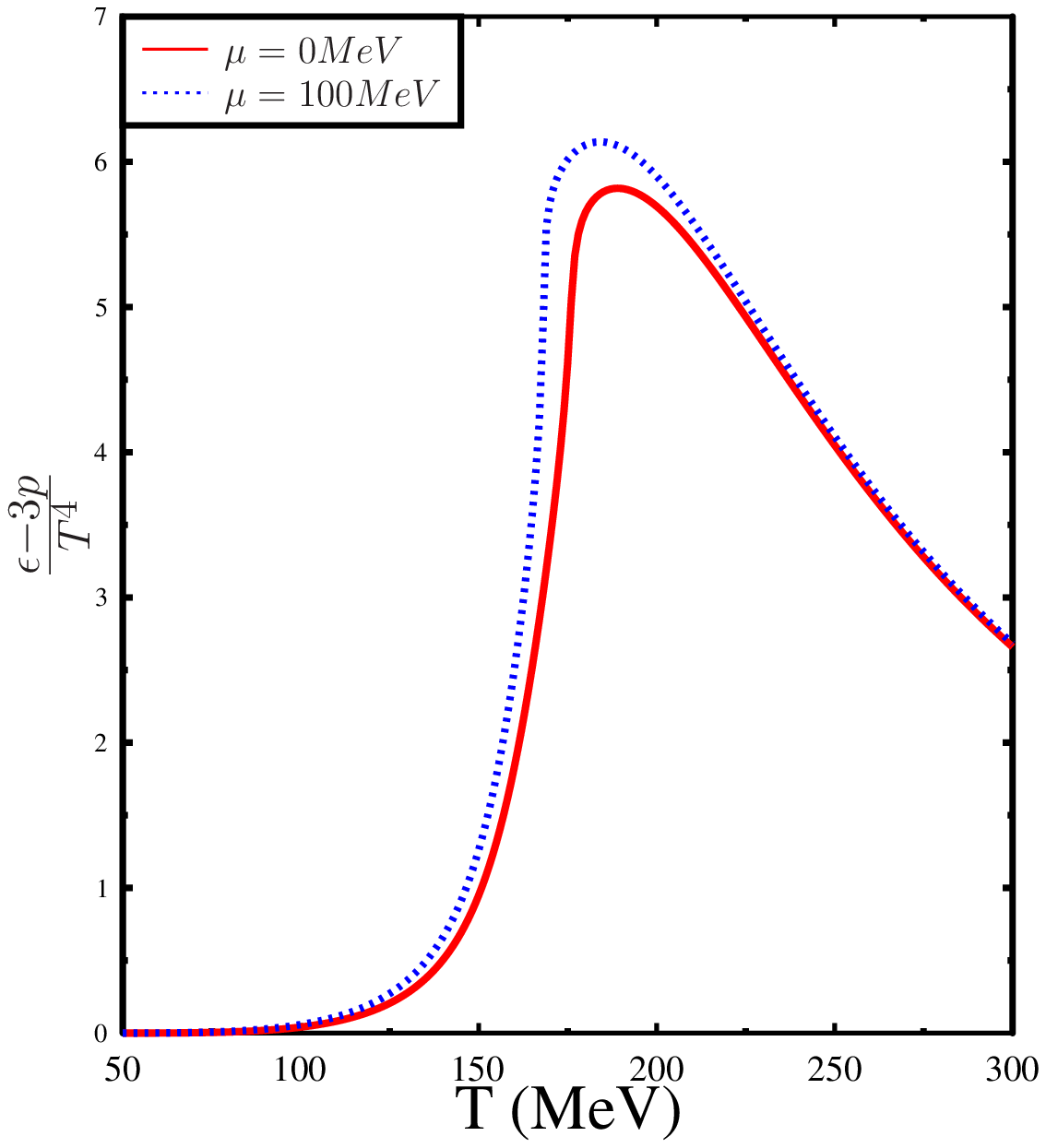}&
\includegraphics[width=9cm,height=9cm]{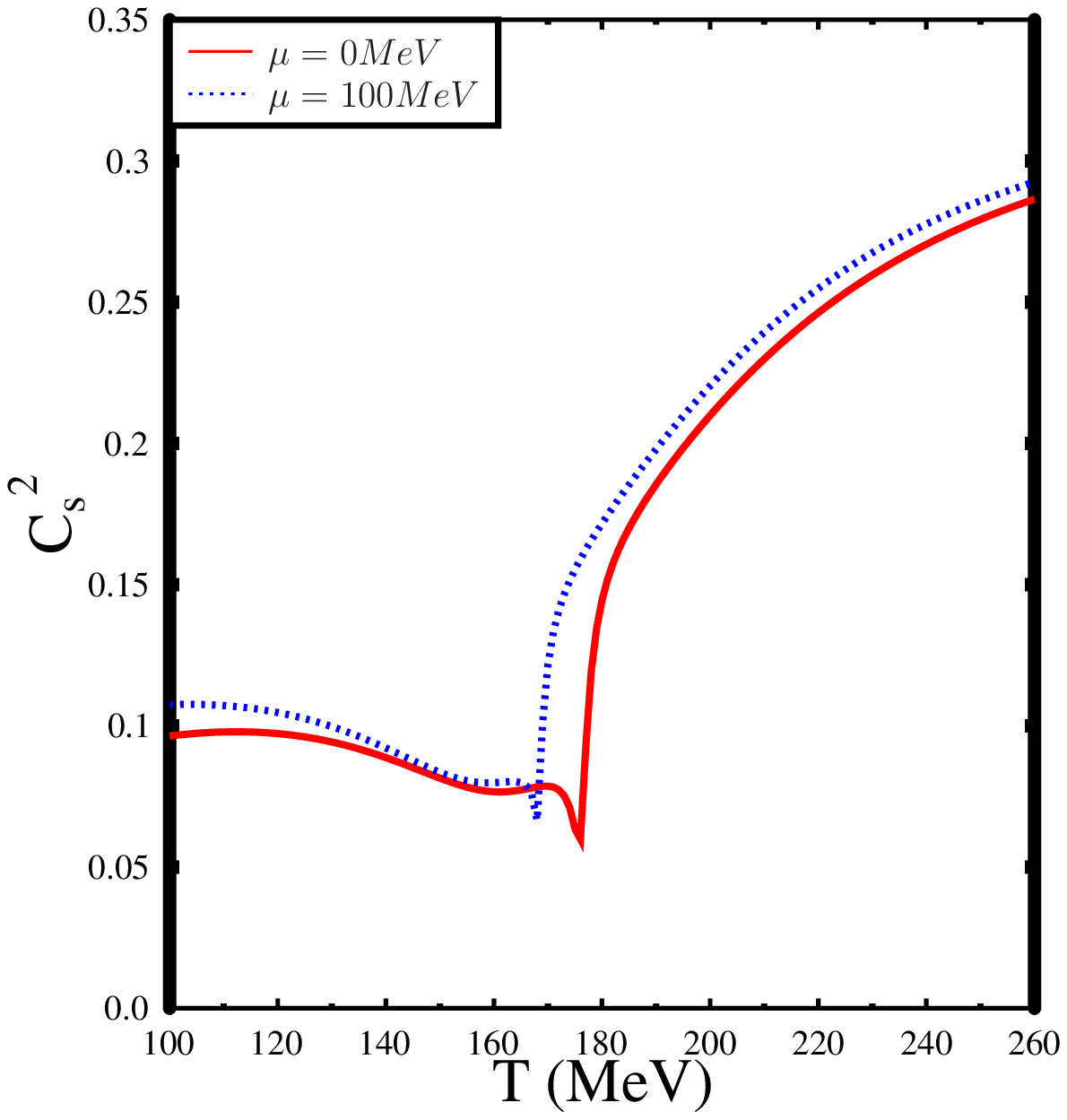}\\
Fig. 2-a & Fig. 2-b
\end{tabular}
\end{center}
\caption{  (Fig 2 a)Temperature dependence of the scaled trace anomaly $\frac{\epsilon-3p}{T^4}$
and (Fig2 b)Temperature dependence of the velocity of sound at constant density. }
\end{figure}


 Next, in Fig 2a, 
we show the dependence of the 
trace anomaly $(\epsilon-3p)/T^4$ on temperature. The conformal symmetry is broken maximally 
at the critical temperature.  Further finite chemical potential
enhances this breaking as it breaks scale symmetry explicitly. As we shall see later this will have its implication on
the bulk viscosity coefficient.

The other thermodynamic quantity that enters into the transport coefficient calculation is the velocity of sound. 
The same at constant density
is defined as
\be
c_s^2= \left(-\frac{\partial P}{\partial \epsilon}\right)_n=
\frac{s\chi_{\mu\mu}-\rho\chi_{\mu T}}{T(\chi_{TT}\chi_{\mu\mu}-\chi_{\mu T}^2)}
\label{sound}
\ee
where, 
$P$,the pressure, is the negative of the thermodynamic potential given in Eq.(\ref{thpot}). Further, 
$s=-\frac{\partial\Omega}{\partial T}$ is the entropy density and  the susceptibilities are defined as 
$\chi_{xy}=-\frac{\partial^2\Omega}{\partial x\partial y}$. This is plotted in Fig. 2b.
The velocity of sound shows a minimum near the crossover temperature. Within the model, at low temperature when the constituent quarks
start contributing to the pressure, their contribution to the energy density is significant compared to their contribution to the pressure
leading to decreasing behavior of velocity of sound till the crossover temperature beyond which it increases as the quarks become 
light and approach the massless limit of $c_s^2=\frac{1}{3}$. Such a dip in the velocity of sound is also observed in lattice simulation \cite{tanmoy}. As we shall observe later this behavior will have important consequences for the
behavior of bulk viscosity as a function of temperature.
We might mention here that such a dip for the sound velocity was not observed for two flavor NJL \cite{deb}. For
in the linear sigma model calculations such a dip was observed only for a large sigma meson mass\cite{purnendu}.
\section{ Transport coefficients in relaxation time approximation}
 Within a quasi-particle approach, a kinetic theory treatment for estimation of transport coefficients can be a reasonable approximation
\cite{quasip}. To solve the relativistic Boltzmann equation, we shall further use the relaxation time approximation
where the particle masses are medium dependent.
Such attempts were made earlier for $\sigma$-model\cite{purnendu} as well as NJL model to compute the shear and bulk viscosity
coefficients. Such an approach was also made to estimate the viscosity coefficients of pure 
gluon matter\cite{quasip}. 
The expressions for the viscosity coefficients were put on a firmer ground by deriving the expressions when there are 
mean fields and medium dependent masses in a quasi particle picture \cite{albright}. The resulting expressions for
 the transport coefficients were manifestly positive definite as they should be.
These expressions were derived explicitly for NJL model in Ref.\cite{deb}.
However, a direct generalization of the  expressions for the transport coefficients in presence of background gluon fields is
not straight forward. The reason being the equilibrium distribution functions
for the quarks and antiquarks as given in Eq.s (\ref{fq}) and (\ref{fqbar}) are not Fermi distribution function. 
To make the discussion simpler
let us consider the case of vanishing baryon chemical potential which we discuss in the following.

Let us first note that the background gluon field couple to quarks through covariant derivative as $D_\mu=\partial_\mu-\delta_{\mu 0}A_0$.
In the Polyakov gauge, the Wilson line is $L$ is in the diagonal representation in the color space and therefore, the background gluon field
act as an imaginary chemical potential for the  colored particles. The corresponding color dependent
equilibrium distribution function for the quarks and the anti-quarks are given by \cite{pisarskijhep}
\be
f_i(E)=\frac{1}{e^{\beta(E-iQ_i)}+1}; \quad \bar f_i(E)=\frac{1}{e^{\beta(E+iQ_i)}+1}
\label{fqi}
\ee

 where, we have written $A_0^{ij}=\frac{1}{g}\delta^{ij}Q^i$, without any summation over the index $i$. As $A_0$ is traceless, $\sum_iQ^i=0$.
The Polyakov loop $\phi$ is thus related to $ Q_i$ as $\phi=\frac{1}{3}\sum_ie^{i\beta Q_i}$. Further, for vanishing
 baryon density, one can choose $\phi$ to be real and parameterize $Q^i=2\pi T(-q,0,q)$ with $q$ as the dimension less condensate variable. The Polyakov loop
variable $\Phi$ is therefore given by 
\be
\phi=\frac{1}{3}(1+2\cos{2\pi q}).
\label{phiq}
\ee

It is easy to check that the the distribution function of Eq.(\ref{fq}) is the color averaged distribution function .i.e
$f_\phi(E)=\frac{1}{3}\sum_i f_i(E)$.

One can write down a Boltzmann kinetic equation for the color dependent the single particle
 distribution function $\f_i$ of Eq.(\ref{fqi}) as

\be
\frac{df_{ia}}{dt}=\frac{P^\mu}{E_a}\partial_\mu f^{ia}-\frac{M}{E_a}\frac{\partial M}{\partial x^i}\frac{\partial f_{ia}}{\partial p^i}=
-C^{ai}(f_{ia}).
\label{boltz}
\ee
 To estimate the transport coefficient, one is interested in small departure from equilibrium and one writes f$_{ia}$=f$_{ia}^0$+f$_{ia}^1$, where f$_{ia}^0$ is the equilibrium distribution function,  f$_{ia}^0=[exp(\beta)u_{\nu}(\zbf x)\mp i\beta Q_i(\zbf x)]^{-1}.$
Within the relaxation time approximation, in the collision term, all the distribution functions are given by the equilibrium 
distribution function except for $f_{ia}$. The collision term then, upto first order in deviation from the equilibrium distribution function,
will be proportional to $f^1_{ia}$ as $C(f_{ia}^0)=0$ by local detailed balance. The collision term is then given by
\be
C(f_{ia})=-\frac{f^1_{ia}}{\tau_{ia}}.
\ee

where, $\tau_{ia}$ is the color dependent relaxation time and is in general a function of energy. One can follow he same procedure
as in Ref.\cite{albright,deb} to calculate e.g. the shear viscosity coefficient $\eta$ and the expression for the same is given by
\be
\eta=\frac{1}{45 T} \sum_{i,a} \int \frac{d\zbf p}{(2\pi)^3} \frac {p_a^4}{E_a^2}\tau_{i,a}(E_a)f_{ia}^0(1- f_{i,a}^0)
\label{eta1}
\ee

In the following we shall replace $\tau_{i,a}(E_a)$ by its color averaged relaxation time $\tau_a(E_a)$ which for $N_c$=3 is given as
\be
\tau_a^{-1}=\frac{1}{3}\sum_i\tau_{ia}(E_a)=\frac{1}{3}\sum_{i,j,k,l}\int  d\Gamma^b d\Gamma^cd\Gamma^d
 W_{ia,jb\to kc,ld}\left[f_{jb}^0(1-f_{kc}^0)(1-f_{ld}^0)\right]
\label{taua1}
\ee

where, $d\Gamma_a=\frac{d\zbf p_a}{(2\pi)^3 2E_a}$, and,
\be
W_{ia,jb\to kc,ld}=(2\pi)^4\delta^4(p_a+p_b-p_c-p_d)|M_{ia,jb\to kc,ld}|^2
\ee
with $|M|^2$ being the corresponding square of the matrix element  for the scattering process.  Now, within the model,
since we do not have dynamical gluons and we consider scattering through meson exchanges, the interactions are are color preserving and,
$W_{ia,jb\to kc,ld}\propto \delta{ik}\delta{jl}$ so that,
\be
\tau_a^{-1}(E_a)=\frac{1}{3}\sum_{i,j}\int  d\Gamma^b d\Gamma^cd\Gamma^d
 W_{ia,jb\to ic,jd}\left[f_{jb}^0(1-f_{ic}^0)(1-f_{jd}^0)\right]
\label{taua2}
\ee

The color sum of the distribution functions  become
\bearr
 {\cal F}(E_a,E_c)&\equiv&
\sum_i f_i^0(E_a)(1-f_i^0(E_c))=3 f_\phi -\frac{1}{D(E_a)D(E_c)}\bigg[3 e^{-3\beta(E_a+E_c)}+3\phi(3\phi-2)e^{-\beta(E_a+E_c)}\nonumber\\
&+&3\phi\left(e^{-\beta(E_a+3E_c)}+e^{-\beta(E_c+3E_a)}\right)
+6\phi^{-2\beta(E_a+E_c)}\nonumber\\
&+&3\phi (3\phi-1)\left(e^{-\beta(E_a+2E_c)}+e^{-\beta(E_c+2E_a)}\right)\bigg],
\label{f2}
\eearr
where, $D(E)$ is the denominator of the Polyakov loop distribution function Eq.(\ref{fphi}),
 $D(E)=1+3 \phi e^{-\beta E}+3 \phi e^{-2\beta E}+3 e^{-3\beta E}$.
Eq.(\ref{taua2}) then reduces to
\be
\tau_a^{-1}(E_a)=\int d\Gamma^b d\Gamma^c d\Gamma^d
 W_{ia,jb\to ic,jd}(1-f_\phi^{0c}){\cal F}(E_b,E_d)
\label{tauquark}
\ee

 The expression for $\eta$, Eq.(\ref{eta1}) using Eq.(\ref{f2}), becomes
\be
\eta=\frac{1}{45 T} \sum_a \int \frac{d\zbf p}{(2\pi)^3} \frac {p_a^4}{E_a^2}\tau_{a}(E_a){\cal F}(E_a,E_a)
\label{eta2}
\ee

One can further approximate the expression for $\eta$ by replacing the distribution functions in Eq.(\ref{eta1}) or equivalently in
Eq.(\ref{f2}) by their color averaged value so that $\eta$ reduces to more familiar expression as
$\eta$ becomes
\be
\eta=\frac{1}{15 T} \sum_a \int \frac{d\zbf p}{(2\pi)^3} \frac {p_a^4}{E_a^2}\tau(E_a)f_a^0(1\pm f_a^0)
\label{eta}
\ee
where, the sum is over all the different species contributing to the viscosity coefficients including the antiparticles,
and, $\tau^a$ is the energy dependent relaxation time given in Eq.(\ref{taua1}) which we shall estimate in the following subsection.
Let us note that while such a replacement of the color averaged distribution function is exact in the Boltzmann limit, the 
leading term for difference between replacing the colored distribution function and their color averaged one in the expression
$\sum_i f_{ia}(1-f_{ia})$ is  proportional to $\phi(\phi-1)e^{-2\beta E}$. This difference is small both below and above the 
critical temperature
while it can be relevant around the critical temperature. We have verified numerically that such a difference does not change
the quantitative values for the transport coefficients except near the critical temperature.

The coefficient of bulk viscosity is given by
\bearr
\zeta &=&\frac{1}{27T} \sum_a\int \frac{d\zbf p}{(2\pi )^3}\frac{\tau ^a}{E_a{}^2}{\cal F}(E_a,E_a)
\bigg[\zbf p^2\left(1-3v_n{}^2\right)-3v_n{}^2\left(M^2-TM \frac{dM}{dT}-{\mu M}
 \frac{{dM}}{d\mu }\right)\nonumber\\
&+&3\left(\frac{\partial P}{\partial n}\right)_{\epsilon }\left(M \frac{dM}{d\mu }-E_at^a\right)\bigg]^2
\label{zeta}
\eearr

The thermal conductivity on the other hand is given by
\be
\lambda=\frac{1}{3}\left(\frac{w}{nT}\right)^2\sum_a\int \frac {d\zbf p}{(2\pi)^3}\frac{\zbf p^2}{3 E_a^2}\tau_a(E_a)
\left(t_a-\frac{nE_a}{w}\right)^2 {\cal F}(E_a,E_a)
\label{lamb}
\ee
In the above, $t_a$ is the quark charge (1/3rd baryonic charge) of the constituent particles i.e. 
$t^a=$+1, -1, 0 for the quarks, the anti-quarks and the  mesons
respectively and $w=\epsilon+p$ is the enthalpy density.

\subsection{Relaxation time estimation- meson scattering}

In the following we shall first
estimate the relaxation times involving meson scattering similar to Ref\cite{purnendu}. The scattering amplitudes
involving meson propagators yield divergent integrals due to poles in the s and u channels. So in these amplitudes, we have taken the limits when the Mandelstam variables are taken to be infinity so that the scattering amplitudes reduce to constants.
The energy dependent relaxation time for the meson species $'a'$ arising from a
scattering process $a,b\to c,d$ is given by,
with $d\Gamma_i=\frac{d\zbf p_i}{2 E_i(\zbf p)(2\pi)^3}$, \cite{deb}
\begin{equation}
\tau(E_a)^{-1}=\sum_b\frac{1}{1+\delta_{ab}}
\int d \Gamma_b 
d\Gamma_{c}d\Gamma_d f_M(E_b)(2\pi)^4\delta^4(p_a+p_b-p_c-p_d)\nonumber\\
|M|^2 (1+f_M(E_c))(1+f_M(E_d)) 
\label{weboson}
\end{equation}
In the above, the summation is over all the particles  except the species $a$ with $a,b$ as the initial state
and $f_M(E_a)$ is the Bose distribution for the meson.

In the limit of constant $|M|^2$,  Eq.(\ref{weboson}), the relaxation time for species 'a'  reduces to
\be
\tau(E_a)^{-1}=\frac{1}{256\pi^3 E_a}\sum_b\int_{m_b}^\infty dE_b \sqrt{E_b^2-m_b^2}f_M(E_b)|M|^2\int_{-1}^{1}
\frac{dx}{1+\delta_{ab}}\frac{\sqrt{\lambda(s,m_a^2,m_b^2)\lambda(s,m_c^2,m_d^2)}}{p_{ab}{s^{3/2}}}.
\label{omegaea}
\ee
In the above, $p_{ab}(s)=1/(2\sqrt s)\sqrt{\lambda(s,m_a^2,m_b^2)}$, and the kinematic function $\lambda(x,y,z)=x^2+y^2+z^2-
2xy-2yz-2zx$. The center of mass energy s is given as
$$s=2E_aE_b\left(1+\frac{m_a^2+m_b^2}{2E_aE_b}-\frac{p_ap_b}{E_aE_b}x\right)$$


\subsection{Relaxation time estimation- quark scattering}
We next consider the quark scattering within the model through the exchange of pion and sigma meson resonances. The approach is similar to 
Ref.s\cite{deb,klevansky,marty} performed within NJL model to estimate the corresponding relaxation time for the quarks and
anti-quarks.  The transition frequency is again given by Eq.(\ref{tauquark}), with the corresponding $W_{ab}$ given as
\be
W_{ia,jb\to ic,jd}^q(s)=\frac{2\sqrt{s(s-4m^2)}}{3(1+\delta_{ab})}\int_{t_{min}}^{0}dt\left(\frac{d\sigma}{dt}|_{ia,jb\to ic,jd}\right)
{\cal F}(\frac{\sqrt s}{2},\frac{\sqrt s}{2})
\label{wsig}
\ee
 where,
\be
\frac{d\sigma}{dt}=\frac{1}{16\pi s (s-4m^2)} \frac{1}{p^2_{ab}} |\bar M|_{ia,jb\to ic,jd}^2
\ee
 For the quark scattering, in the present case for 
two flavors we consider the following twelve possible scattering processes:
$u\bar u\rightarrow u\bar u,\quad u\bar d\rightarrow u\bar d,\quad u\bar u\rightarrow d\bar d,$
$u u\rightarrow u u,\quad u d\rightarrow u d,\quad \bar u\bar u\rightarrow\bar u \bar u,$
$\bar u\bar d\rightarrow \bar u\bar d,\quad d\bar d\rightarrow d\bar d,\quad d\bar d\rightarrow u\bar u,$
$d\bar u\rightarrow d\bar u,\quad d d\rightarrow d d,\quad \bar d\bar d\rightarrow \bar d\bar d,$
One can use $i$-spin symmetry, charge conjugation symmetry and crossing symmetry to relate the matrix element
square for the above 12 processes to get them related to one another and one has to evaluate only two independent matrix elements
to evaluate all the 12 processes. We  choose these, as in Ref. \cite{klevansky}, to be the processes
$u\bar u\rightarrow u\bar u$ and $u\bar d\rightarrow u\bar d$ and use the symmetry conditions to calculate the rest.
The square of the  matrix elements for these two processes are given explicitly in  Refs\cite{deb,klevansky}
in terms of Mandelstam variables and the meson propagators. In the present model, the
meson propagators $D_{a}(\sqrt{s},0)$, ($a=\sigma,\zbf \pi$) are given by
\be
D_{a}(\sqrt{s},\zbf 0)=\frac{i}{s-M_a^2-i Im\Pi_{M_a}(\sqrt{s},\zbf 0)}
\label{propmeson}
\ee
In the above, the masses of the mesons are given by M$_a$'s which are medium dependent masses for mesons determined by the curvature of 
the thermodynamic potential.
Further, in Eq.(\ref{propmeson}), $Im\Pi(\sqrt{s},0)$ which is related to the width of the resonance as 
$\Gamma_a=Im\Pi_a/M_a$ is given as
\cite{klevansky}
\be
Im\Pi_a(\omega,\zbf 0)=\theta(\omega^2-4 m^2)\frac{N_cN_f}{8\pi\omega}\left(\omega^2-\e_a^2\right)\sqrt{\omega^2-4m^2}
\left(1-f_-(\omega)-f_+(\omega)\right)
\ee
with $\e_a=0$ for pions and $\e_a=2m$ for sigma mesons.

\subsection{Quark pion scattering and relaxation time}

Next, we compute the contribution of quark meson scattering to the relaxation times for both mesons as well as
quarks. In the following we consider the quark pion scattering only as the sigma meson contribution is negligible. The Lorentz invariant scattering matrix element
can be written as $\bar U(p_2)T_{ba}U(p_1)$, with $\bar UU=2m_q$ and with $p_1,p_2$ denoting the
initial and final the quark momenta respectively and $q_1,q_2$, being the momenta of the pions.
\be
T_{ba}=\delta_{ba}\frac{1}{2}(q_1+q_2)^\mu\gamma_\mu (\delta_{ab}B^{(+)}+i\epsilon_{abc}\tau_c B^{(-)})
\ee
 where,
\be
B^{(+)}=g^2\left(\frac{1}{u-m_q^2}-\frac{1}{s-m_q^2}\right),
\label{bplus}
\ee
and
\be
B^{(-)}=-g^2\left(\frac{1}{u-m_q^2}+\frac{1}{s-m_q^2}\right).
\label{bminus}
\ee

Averaging over the spin and isospin factors, the matrix element square for the quark pion scattering is given by
\be
|\bar M|^2=\frac{g_\sigma ^4}{6}\left((s-u)^2-t(t-4m_\pi^2)\right)\left(3B_+^2+2 B_-^2\right)
\label{modm2qpi}
\ee
The contribution to quark relaxation time from the quark pion scattering is given by, Eq,E$_\pi$ being the center of mass energies of outgoing quark and pion respectively.
\be
\tau_q(E_q)|_{q\pi}=\frac{1}{32\pi E_q}\int d\zbf \pi_b f_\pi(E_b) \frac{1}{\sqrt{s}p_0}\int dt |\bar M_{q-\pi}|^2
(1-f_\phi(Eq))(1+f_\pi(E_\pi))
\ee
In the above, $p_0^2= (s+m_q^2-m_\pi^2)^2/(4s)-m_q^2$.
On the other hand, the contribution to the pion relaxation time arising from quark pion scatterings
is given by
\be
\tau_\pi(E_\pi)|_{q\pi}=\frac{1}{96\pi E_\pi}\int d\zbf \pi_b  \frac{1}{\sqrt{s}p_0}\int dt |\bar M_{q-\pi}|^2
(1+f_\pi(\sqrt s/2)){\cal F}(E_b,E_q)
\ee

Let us note that there are poles in the u channel in the quark pion 
scattering term beyond the critical temperature
when the pion mass become larger than the quark mass. However, this is taken care of once we include 
the imaginary part of the quark self energy in the propagators for the
quarks in the calculation of the amplitude in Eq.s(\ref{bplus})-(\ref{bminus}).
\cite{lang}

\section {Results}
\begin{figure}
 \begin{center}
 \begin{tabular}{c c}
 \includegraphics[width=9cm,height=9cm]{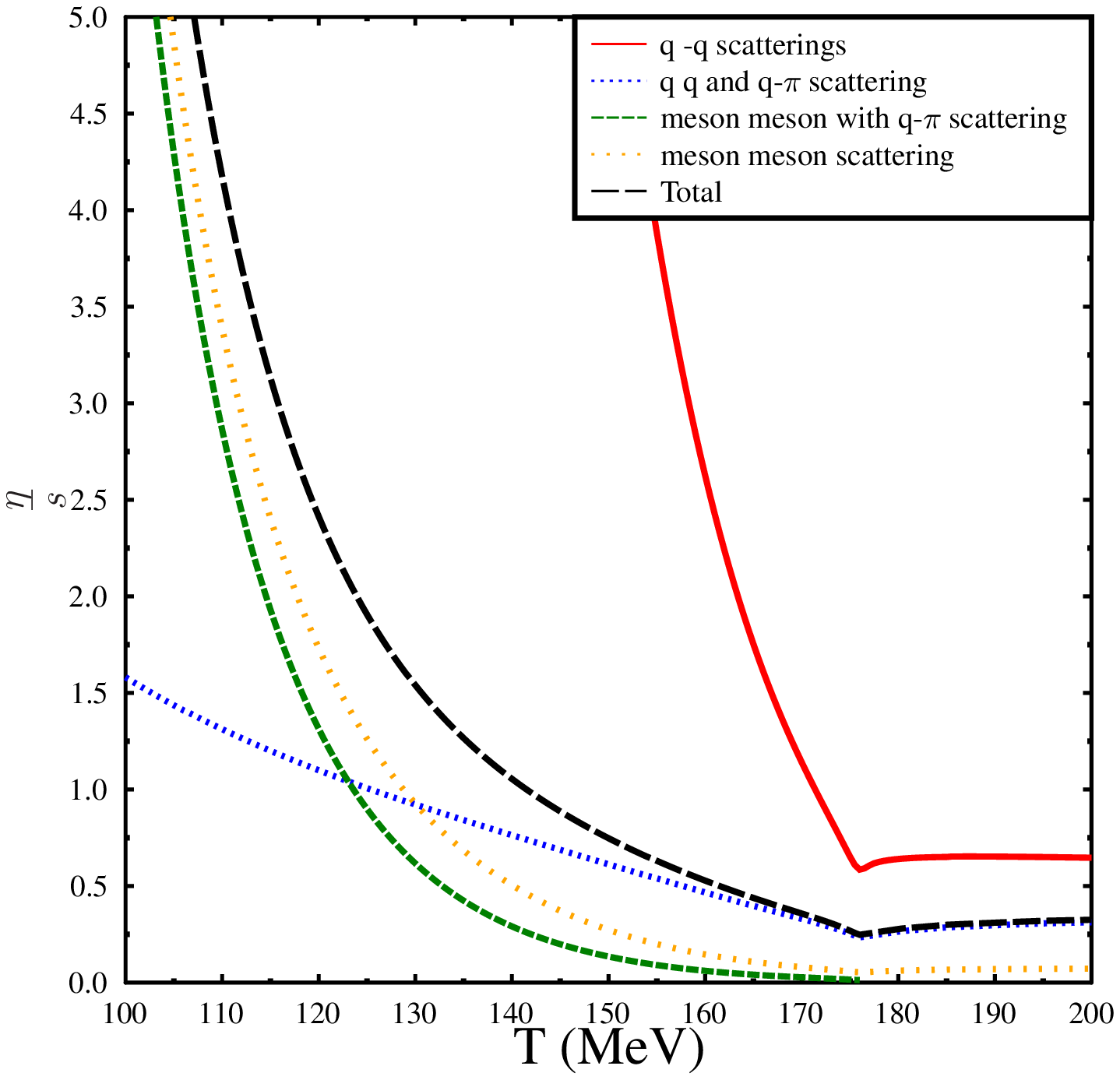}&
 \includegraphics[width=9cm,height=9cm]{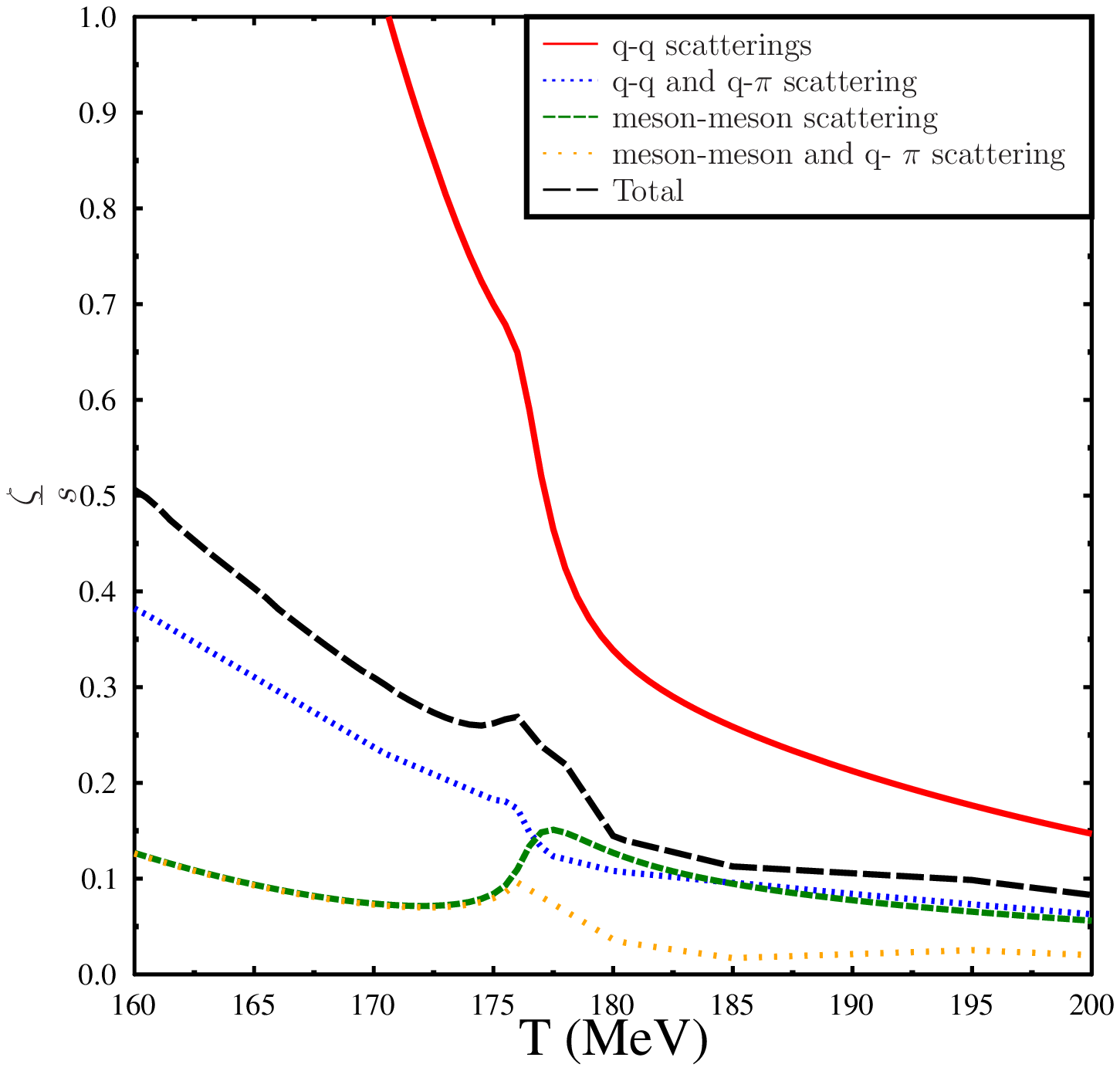}\\
 Fig. 3-a & Fig. 3-b
 \end{tabular}
 \end{center}
 \caption { Different contributions for specific viscosity coefficients.
$\eta/s$ is shown in the left while $\zeta/s$ is shown on the right.
In both the figures, contributions from the quarks arising from
quark quark scattering ( red solid line) and including quark meson scattering
 time for  (blue dotted line) is shown as a function of temperature.
 The contribution of the mesons due to meson meson scattering
 (green dashed curve) and including meson quark scattering (orange short dashed curve)
is also shown. The total contribution from the quarks and mesons are
is shown by the black long dashed curve.
All the curves correspond to $\mu=0$ case.}
 \label{fig8}
 \end{figure}

  We now discuss about the contribution of different scatterings to the relaxation time and hence their
contribution to the  
specific shear viscosity $\eta/s$. 
This is  shown in Fig.\ref{fig8}a for
vanishing chemical potential. The contribution from the mesons to the shear viscosity is arising from the meson
-meson scattering only is shown by the green dashed curve while the effect of including the meson quark scattering 
in the relaxation time estimation is shown by the orange dotted curve. Similarly the quark contribution to 
this ratio $\eta/s$ with a relaxation time  arising from quark quark scattering is shown by the red solid line while the 
quark contribution to the viscosity with a relaxation time estimated including 
 the quark pion scattering is shown by the blue dotted line. This also demonstrates the importance of the 
scattering of quarks and mesons to the total viscosity coefficient. The total contributions from both the quarks 
and mesons is shown as the black dashed curve in Fig.\ref{fig8}a. Considering the contribution from the mesons, as may be see, including only the meson meson scattering the specific shear viscosity shows a minimum at the critical temperature 
with a numerical value $\eta/s\sim 0.053$which is lower than the KSS bound of $1/4\pi$. Inclusion of
meson scattering with quarks however increase this value. With regards to the quark contributions (the red solid line in Fig
\ref{fig8}a), the dominant contribution
here comes from quark antiquark scattering through s-channel meson exchange. The masse of the $\sigma$-meson decreases with temperature becoming a minimum at $T_c$ leading to an enhancement of the cross section. Beyond $T_c$, the meson masses increase
leading to a decrease of the cross section. This leads to a minimum of the relaxation time and hence the shear
viscosity arising from quark quark scattering. Further, the effect of Polyakov loop lies in suppressing the
cross section below the critical temperature as compared to e.g. Nambu JoanaLasinio models\cite{deb} leading to s sharp increase of
the relaxation time and hence the viscosity below the $T_c$. However, when the quark meson scattering effects are included for
one would have expected this contribution from the quark meson scattering would be suppressed due to increasing meson masses beyond $T_c$. However, beyond the critical temperature, there are poles in the u-channels for $q-\pi$ scattering as the $M_\pi$ become larger than the quark masses. This is however, regulated by the finite width of the quarks. None the less, the contribution of 
$q-\pi$ scatterings to the quark relaxation time remains non-negligible beyond $T_c$. The total contribution to the 
ratio $\eta/s$ is shown as black dashed curve. Clearly, the meson contributions to this ratio dominate at temperatures
below $T_c$ while, the quark contribution dominate this ratio above $T_c$ as one would expect.

In a similar manner,  various contributions to the specific bulk 
viscosity ($\zeta/s$) coefficient is shown in Fig\ref{fig8}b. 
The notation regarding different contributions to $\zeta/s$ is same
as in Fig.\ref{fig8}a. As may be noted, while a peak structure is seen for the contribution
arising from meson-meson scattering ( green dashed curve) at the critical temperature, such a peak is
somewhat reduced when meson quark scattering is included. Similarly, for the $q-q$ scattering contribution, no such peak structure is seen and such a result is similar to what is seen in NJL models \cite{deb}. On the other hand, when one includes the contribution of quark meson scatterings, a peak structure is seen for $\zeta/s$. The total effect is shown as the black dashed curve which shows a small peak structure near $T_c$.


\begin{wrapfigure}{r}{0.5\textwidth}
  \vspace{-20pt}
  \begin{center}
    \includegraphics[width=0.48\textwidth]{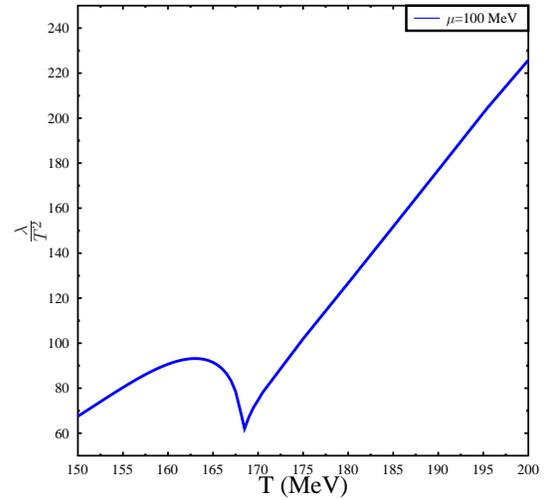}
  \end{center}
  \vspace{-20pt}
  \caption{Thermal conductivity in units of $T^2$ as a function of temperature for $\mu=100$MeV. }
  \vspace{-10pt}
  \label{fig12}
\end{wrapfigure}

In Fig.\ref{fig12}, we have shown the results for thermal conductivity. We have plotted here the dimensionless quantity $\lambda/T^2$ as a function of
temperature for $\mu=100 $MeV.  As is well known, thermal conductivity for relativistic particles 
actually diverges for $\mu=0$ and
the heat conduction vanishes. However, for situations where e.g. pion number is
conserved heat conductivity can be sustained by pions which 
themselves have zero baryon number. What we have shown in
 Fig.(\ref{fig12}) is the thermal conductivity arising only from quark scattering.
Similar to the behavior of relaxation time, the specific thermal conductivity has a minimum at $T_c$.
This behavior of having a minimum at $T_c$ is
similar to Ref.\cite{deb} for NJL
model. The sharp rise of $\lambda/{ T}^2$ can be understood by 
performing  a dimensional argument to show that at very high temperature when 
chiral symmetry is is restored the integral increases 
as $ T^3$ while the prefactor $w/(nT)$
grows as $ T^2$ for small chemical potentials. Apart from this 
kinematic consideration, as the integrand further is multiplied
by $\tau(E)$ which itself is an increasing function of temperature 
beyond $T_c$, leads to the sharp rise of the
ratio $\lambda/T^2$ beyond the critical temperature. Below, the critical 
temperature, however, the ratio decreases which is in contrast to NJL results
of Ref.\cite{deb}. The reason is two fold. Firstly the magnitude of relaxation time decrease 
when quark meson scattering is included as compared to quark quark scattering.
This apart, in the integrand,the distribution functions are suppressed by Polyakov loops as compared to NJL model.
      
\section*{Summary}

Transport coefficients of hot and dense matter are important inputs
for the hydrodynamic evolution of the plasma that is produced following a heavy ion collision. In the present study, we have  estimated coefficients
taking into account the the non-perturbative effects related to chiral
 symmetry breaking and the confinement properties of strong interaction physics within
an effective model, the Polyakov loop extended quark meson coupling model.
These coefficients are estimated using relaxation time approximation
for the solutions of the Boltzmann kinetic equation.

We first calculated the medium dependent masses of the mesons and quarks within
a mean field approximation. The contribution of the mesons to the 
transport coefficients has been calculated through estimating 
the relaxation time for the mesons arising both from meson meson scattering 
and meson quark scattering. The 
contribution to the transport coefficients arises mostly from the meson 
scatterings at temperatures below the critical temperature
 while above the critical temperature the contributions arising 
from the quark scatterings become dominant. In particular, 
quark meson scattering contribute significantly to the
relaxation time for the quarks both below and above the critical temperature. 
The quark pion scattering above the critical temperature gives significant 
contribution due to the pole structure of the corresponding 
scattering amplitude. 

In general, the effect of Polyakov loops lies in suppressing the quark 
contribution below the critical temperature. This leads to, in particular,
 the suppression of thermal conductivity at lower temperature arising 
from quark scattering.
The effect of Polyakov loop also is significant near and above the critical
temperature. Indeed, both the quark masses as well as Polyakov loop
order parameter remain significantly different from their
asymptotic values  near the critical temperature.
It will be interesting to examine the consequences of such non-perturbative 
features on the
transport coefficients of heavy quarks as well as on the 
collective modes of QGP
above and near the critical temperature. Some of these works are in progress and will be reported elsewhere.


\def\berera{A. Bstero-Gil, A. Berera and R. Ramos, JCAP1107, 030 (2011).}
\def\heinzrev{U. Heinz and R. Snellings, Annu. Rev. Nucl. Part. Sci. 63, 123-151, 2013}
\def\pethick{H. Heiselberg and C. Pethick,{\PRD{48}{2916}{1993}}.}
\def\exptstar{J. Adams,{\em et al}(STAR),{\NPA{757}{102}{2005}}.}

\def\karschkharzeev{F. Karsch, D. Kharzeev, and K. Tuchin, Phys. Lett. B
663, 217 (2008).}
\def\kharzeevtuchin{D. Kharzeev, and K. Tuchin,JHEP 0808,031, (2008).}
\def\nicolaprl{	
D. Fernandez-Fraile, A. Gomez Nicola, {\PRL{102}{121601}{2009}}.}
\def\joglekar{J.C. Collins, A. Duncan, S.D. Joglekar, Phys. Rev. D 16, 
438 (1977).}
\def\blaschke{J. Jankowski, D. Blaschke, M.Spalinski, Phys.Rev.D 87, 105018^M
(2013). }
\def\gorenstein{M. Gorenstein, M. Hauer, O. Moroz, Phys.Rev.C 77,024911 (2008)}
\def\bugaev{K. Bugaev et al, Eur.Phys.J. A 49, 30 (2013)}
\def\cpsingh{S.K. Tiwari, P.K. Srivastava, C.P. Singh, Phys.Rev. C 85,
014908 (2012) }
\def\chen{J.-W. Chen, Y-H. Li, Y.-F. Liu, and E. Nakano, Phys. Rev. D 76,
114011 (2007)}
\def\chennakano{J.-W. Chen, and E. Nakano, Phys. Lett. B 647, 371 (2007)}
\def\itakura{K. Itakura, O. Morimatsu, and H. Otomo, Phys. Rev. D 77, 014014
(2008)}
\def\cleymans{J. Cleymans, H. Oeschler, K. Redlich, and S. Wheaton, Phys.
Rev. C 73, 034905 (2006)}
\def\worku{J. Cleymans and D. Worku, Mod. Phys. Lett. A26,1197,(2011).}
\def\guptagod{S. Chatterjee, R. M. Godbole and S. Gupta, {\PRC{81}{044907}{2010}}.}
\def\Noronha{Noronha-Hostler J, Noronha J and Greiner C 2012 Phys. Rev. C 86 024913}
\def\tanmoy{A. Bazavov {\it etal}, e-print:arXiv:1407.6387.}
\def\cavitation{K. Rajagopal and N. Trupuraneni, JHEP1003, 018(2010);
 J. Bhatt, H. Mishra and V. Sreekanth, JHEP 1011, 106,(2010);{\it ibid} Phys. Lett. B704, 486 (2011); {\it ibid} Nucl. Phys. A875, 181(2012).}
\def\borsonyi{S. Borsonyi{\it etal}, JHEP1011, 077 (2010).}
\def\borsonyimu{S. Borsonyi{\it etal}, JHEP1208, 053 (2012).}
\def\dobadoch{A. Dobado,F.J.Llane-Estrada amd J. Torres Rincon, 
 J. Bhatt, H. Mishra and V. Sreekanth, JHEP 1011, 106,(2010);{\it ibid} Phys. Lett. B704, 486 (2011); {\it ibid} Nucl. Phys. A875, 181(2012).}
\def\borsonyi{S. Borsonyi{\it etal}, JHEP1011, 077 (2010).}
\def\borsonyimu{S. Borsonyi{\it etal}, JHEP1208, 053 (2012).}
\def\dobadoch{A. Dobado,F.J.Llane-Estrada amd J. Torres Rincon, 
{\PLB{702}{43}{2011}}.}
\def\dobadoshear{A. Dobado,F.J.Llane-Estrada amd J. Torres Rincon, 
{\PRD{79}{055207}{2009}}.}
\def\sasakiqp{C. Sasaki and K.Redlich,{\PRC{79}{055207}{2009}}.}
\def\sasakinjl{C. Sasaki and K.Redlich,{\NPA{832}{62}{2010}}.}
\def\ellislet{I.A. Shushpanov, J. Kapusta and P.J. Ellis,{\PRC{59}{2931}{1999}}
; P.J. Ellis, J.I. Kapusta, H.-B. Tang,{\PLB{443}{63}{1998}}.}
\def\prakashwiranata{Anton Wiranata and Madappa Prakash,
{\PRC{85},{054908}{2012}}.}
\def\purnendu{P. Chakraborty and J.I. Kapusta {\PRC{83}{014906}{2011}}.}
\def\greco{S.Plumari,A. Paglisi,F. Scardina and V. Greco,{\PRC{83}{054902}{2012}a.}}
\def\bes{H. Caines, arXiv:0906.0305 [nucl-ex], 2009.}
\def\greinerprl{J. Noronha-Hostler,J. Noronha and C. Greiner,
{\PRL{103}{172302}{2009}}.}
\def\greinerprc{J. Noronha-Hostler,J. Noronha and C. Greiner
, {\PRC{86}{024913}{2012}}.}
\def\igorgreiner{J. Noronha-Hostler, C. Greiner and I. Shovkovy,
, {\PRL{100}{252301}{2008}}.}
\def\majumdermueller{A. Majumder and B. Mueller, {\PRL{105}{252002}{2010}}.}
\def\leonidov{ A. V. Leonidov and P. V. Ruuskanen, {\EPJC{4}{519}{1998}}.}
\def\cbm{ B. Friman, C.H. Ohne, J. Knoll, S. Leupold, J. Randrup, R. Rapp, P. Senger (Eds.), Lect. Notes Phys., vol. 814,
2011.}
\def\nica {A.N. Sissakian, A.S. Sorin, J. Phys. G 36 (2009) 064069.}
\def\nakano{J.W. Chen,Y.H. Li, Y.F. Liu and E. Nakano,
 {\PRD{76}{114011}{2007}}.}
\def\wang{M.Wang,Y. Jiang, B. Wang, W. Sun and H. Zong, Mod. Phys. lett.
{\bf A76}, 1797,(2011).}
\def\agasian{N.O. Agasian, JETP Lett. 95, 171, (2012), arXiv:1109.5849.}
\def\Hagedorn{R. Hagedorn and J. Rafelski,{\PLB{97}{136}{1980}}.}
\def\kapustaolive{J.I. Kapusta and K. A. Olive, {\NPA{408}{478}{1983}}.}
\def\hrgexp{P. Braunmunzinger, J. Stachel, J.P. Wessels and N. Xu,
{\PLB{365}{1}{1996}}; G.D. Yen and M.I. Gorenstein, {\PRC{59}{2788}{1999}};
F. Becattini, J. Cleymans, A. Keranen, E. suhonen and K. Redlich, 
{\PRC{64}{024901}{2001}}.}
\def\rischkegorenstein{.D.H. Rischke, M.I. Gorenstein, H. Stoecker and
W. Greiner, Z.Phys. C {\bf 51}, 485 (1991).}
\def\hmnjl{Amruta Mishra and Hiranmaya Mishra, {\PRD{74}{054024}{2006}}.}
\def\pdgb{C. Amseler {\it et al}, {\PLB{667}{1}{2008}}.}
\def\shuryak{E.V. Shuryak, Yad. Fiz. {\bf 16},395, (1972).}
\def\leupold{S. Leupold, J. Phys. G{\bf32},2199,(2006)}
\def\peter{A. Andronic, P. Braun-Munzinger , J. Stachel and M. Winn,
{\PLB{718}{80}{2012}}}
\def\blum{M. Blum, B. Kamfer, R. Schluze, D. Seipt and U. Heinz,{\PRC{76}{034901}{2007}}.}
\def\jaminplb{M. Jamin, {\PLB{538}{71}{2002}}.}
\def\ghosh{Sabyasachi Ghosh, {\PRC{90}{025202}{2014}}; International Journal Of Modern Physics {\bf A29}, 145005,2014.}
\def\csernai{L.P. Csernai, J.I. Kapusta and L.D. McLerran,{\PRL{97}{152303}{2006}}.}
\def\hagedorn{R. Hagedorn, Nuovo Cim. Suppl. 3,147 (1965); Nuovo Sim. A56,1027 (1968).}
\def\torieri{G. Torrieri and I. Mishustin,{\PRC{77}{034903}{2008}}.}
\def\fernandez{D. Fernandiz-Fraile and A.G. Nicola,{\PRL{102}{121601}{2009}}.}
\def\caron{S.Caron,{\PRD{79}{125009}{2009}}.}
\def\latticemeyer{H.B. Meyer,{\PRL{100}{162001}{2008}}.}
\def\romatschke{P.Romatscke and D.T. Son,{\PRD{80}{065021}{2009}}.}
\def\moore{G.D. Mooore and O. Sarem, J. High Energy Phys. JHEP0809(2008)015.}
\def\dobado{A.Dobado and J. M. Torres-Rincon {\PRD{86}{074021}{2012}}.}
\def\monai{A. Monnai and T. Hirano, Phys. Rev. C 80, 054906 (2009).}
\def\kodama{G. S. Denicol, T. Kodama, T. Koide, and P. Mota, Phys. Rev. C 80, 064901 (2009).}
\def\heinz{H. Song and U. Heinz, Phys. Rev. C 81, 024905 (2010).}
\def\bozek{P. Bozek, Phys. Rev. C 81, 034909 (2010).}
\def\schaferdus{K. Dusling and T. Schafer,  ̈ Phys. Rev. C 85, 044909 (2012).}
\def\noronhahydro{J. Noronha-Hostler, G. S. Denicol, J. Noronha, R. P. G. Andrade,
and F. Grassi, Phys. Rev. C 88, 044916 (2013); J. Noronha-Hostler, J. Noronha and F. Grassi,
Phys. Rev. C 90, 034907 (2014)}
\def\rosegale{J. B. Rose, J. F. Paquet, G. S. Denicol, M. Luzum, B. Schenke, S. Jeon and C. Gale, 
{\NPA{931}{926}{2014}}.}
\def\bassprl{ N.~Demir and S.~A.~Bass,
Phys. Rev. Lett. {\bf 102}, 172302 (2009).}
\def\phsdbrat{V.~Ozvenchuk, O.~Linnyk, M.~I.~Gorenstein, E.~L.~Bratkovskaya and W.~Cassing, Phys. Rev. C {\bf 87},  064903 (2013).}
\def\florkowski{W. Broniowski and W. Florkowski, {\PLB{673}{142}{2009}}.}
\def\albright{M. Albright and J.I. Kapusta,{\PRC{93}{014903}{2016}}.}
\def\hirano{P. Romatschke and U. Romatschke, Phys. Rev. Lett.{\bf 99},172301, (2007); T. Hirano and
 M. Gyulassy, Nucl. Phys. {\bf A 769}, 71, (2006).}
\def\daniel{P. Danielewicz, M. Gyulassy, {\PRD{31}{53}{1985}}.}
\def\kss{P. Kovtun, D.T. Son and A.O. Starinets, Phys. Rev. Lett.{\bf 94},
 111601, (2005).}
\def\schenke{C.Gale, S. Jeon and B. Schenke, International Journal of Modern Physics A {\bf 28}, 134011,(2013).}
\def\denicolhydro{G.S. Denicol, H. Niemi, E. Molnar and D.H. Rischke,{\PRD{85}{114047}{2012}}.}
\def\denicolpre{M. Greif, F. Reining, I. Bouras , G.S. Denicol, Z. Xu and  C. Greiner, Phys.Rev. {\bf E87} ,033019(2013).}
\def\denicolheat{G.S. Denicol, H. Niemi, I. Bouras E. Molnar , Z. Xu , D.H. Rischke, C. Greiner ,{\PRD{89}{074005}{2014}}.}
\def\rincon{J.I. Kapusta and J.M. Torres-Rincon,{\PRC{86}{054911}{2012}}.}
\def\rinconprd{J.I. Kapusta and J.M. Torres-Rincon,{\PRC{86}{054911}{2012}}.}
\def\ghoshthermal{Sabyasachi Ghosh, International Journal of Modern Physics {\bf E24},1550058,2015. }
\def\kubo{R. Kubo,J. Phys. Soc. Jpn. {\bf 12},570,(1957).}
\def\klevansky{P. Zhuang,J. Hufner, S.P. Klevansky and L. Neise,{\PRD{51}{3728}{1995}}. } 
\def\klevanskynpa{P. Rehberg, S.P. Klevansky and ,J. Hufner,{\NPA{608}{356}{1996}}. } 
\def\transqcd{P. Arnold,G.D. Moore and L.G. Yaffe, JHEP, 11, 2000, 001; ibid, JHEP 01 (2003) 030; ibid, JHEP 05 (2003) 051}
\def\quasip{M. Bluhm, B. Kamfer and K. Redlich,{\PRC{79}{055207}{2009}}, ibid,{\PRC{84}{025201}{2011}}.}
\def\blumredlich{M. Bluhm, B. Kamfer and K. Redlich,{\PRC{84}{025201}{2011}}.}
\def\marty{R. Marty, E. Bratkovskaya, W. Cassing, J. Aichelin and H . Berrehrah, {\PRC{88}{045204}{2013}}.}
\def\voskresenskynpaa{A.S. Khvorostukhin, V.D. Toneev and D.N. Voskresensky,{\NPA{915}{158}{2013}}.}
\def\voskresensky{A.S. Khvorostukhin, V.D. Toneev and D.N. Voskresensky,
{\NPA{845}{106}{2010}}.}
\def\gavin{Sean Gavin,{\NPA{435}{826}{1985}}.}
\def\hosoya{A. Hosoya and K. Kajantie ,{\NPB{250}{666}{1985}}.}
\def\degroot{ S.R. deGroot, W.A. van Leeuwen and Ch. G. van Weert,{\it Relativistic Kinetic Theory: Principles and Applications( North-Holland, Amsterdam, 1980)}.}
\def\lang{R. Lang, N. Kaiser, W. Weise, Eur. Phys. {\bf A48}, 109, 2012.}
\def\langweise{R. Lang, N. Kaiser, W. Weise, Eur. Phys. {\bf A50 }, 63, 2014.}
\def\ghoshkrein{Sabyasachi Ghosh, Thiago C. Peixoto, Victor Roy, Fernando E. Serna, Gastão Krein e-Print: arXiv:1507.08798 [nucl-th], (2015).}
\def\mpiexpt{K. Hagiwara {\em et al},{\PRD{66}{010001}{2002}}.}
\def\fpiexpt{B. Holostein,{\PLB{244}{83}{1990}}.}
\def\condsum{H.G. Dosch and S. Narrison,{\PLB{417}{173}{1998}}.}
\def\condlat{L. Giusti,F. Rapuano, M. Talevi and A. Vladikas,{\NPB{538}{249}{1999}}.}
\def\matiello{S. Matiello, arXiv:1210.1038[hep-ph].}
\def\nicola{ D. Fernandiz-Fraile and A. Gomez Nicola, {\EPJC{62}{37}{2009}}.}
\def\nam{ S.  Nam, Mod. Phys. Lett. A 30, 1550054,2015.}
\def\fukutome{ M. Iwasaki and T. Fukutome, J. Phys. G36, 115012, 2009.}
\def\sourav{ S. Mitra and S. Sarkar,{\PRD{89}{054013}{2014}}.}
\def\sabyath{S. Ghosh, Int.J.Mod.Phys. E24 (2015) 07, 1550058}
\def\prakash{ M. Prakash, M. Prakash, , R. Venugopalan and G. Welke,{\PR{227}{321}{1993}}.}
\def\buballarev{M. Buballa,{\PR{407}{205}{2005}}.}
\def\heckmann{K. Heckmann, M. Buballa and J. Wambach ,{\EPJA{48}{142}{2012}}}.
\def\quack{E. Quack, P. Zhuang, Y. Kalinovsky, S.P. Klevansky and
J. Hufner,{\PLB{348}{1}{1995}}}.
\def\zhuang{P. Zhuang,J. Hufner, S.P. Klevansky {\NPA{576}{525}{1994}}. } 
\def\gondolo{J. Edsjo, and P. Gondolo, {\PRD{56}{1879}{1997}}. } 
\def\goity{J. I. Goity, and H. Leutwyler, {\PLB{228}{517}{1989}}. } 
\def\page12{D. Page and S. Reddy, (2012), arXiv:1201.5602 [nucl-
th].}
\def\yakovlev{D. G. Yakovlev, A. D. Kaminker, O. Y. Gnedin, and P. Haensel, Phys. Rept. 354, 1 (2001),
 arXiv:astro- ph/0012122 [astro-ph].}
\def\kaminker{D. G. Yakovlev, O. Y. Gnedin, A. D. Kaminker, K. P.
Levenfish, and A. Y. Potekhin, Adv. Space Res. 33,
523 (2003).}
\def\rmode{N. Andersson, Astrophys. J. 502, 708 (1998), arXiv:gr-
qc/9706075 [gr-qc];
N. Andersson and K. D. Kokkotas, Mon. Not. Roy. As-
tron. Soc. 299, 1059 (1998), arXiv:gr-qc/9711088 [gr-
qc].}
\def\jharmode{ T.K. Jha, H. Mishra, V. Sreekanth,{\PRC{82}{025803}{2010}}.}
\def\kapustabk{J. I. Kapusta,{\underline Relativistic Nuclear Collisions}, Landolt-Bornstein new Series, Vol I/23,
Ed. R. Stock (Springer Verlag, Berlin Heidelberg 2010).}
\def\deb{Paramita Deb, Guru Prakash Kadam, Hiranmaya Mishra, {\PRD{94}{094002}{2016}}.}
\def\Gyulassylarry2005{M. Gyulassy and L. McLerran, {\NPA{750}{30}{2005}}.}
\def\luzumromatschke{M. Luzum and P. Romatschke, {\PRL{103}{262302}{2009}}.}
\def\dirkprl{H. Niemi, G.S. Denicol, P. Huovienen, E. Molnar and D.H. Rischke,{\PRL{106}{212302}{2011}}.}
\def\souravsuk{S. Mitra and S. Sarkar, {\PRD{87}{094026}{2013}, S. Mitra, S. Gangopadhyaya, S. Sarkar, {\PRD{91}{094012}{2015}}.}}
\def\kapustalarryprl{L. P. Csernai, J. I. Kapusta and L. D. McLerran, {\PRL{97}{152303}{2006}}.}
\def\pracheta{Pracheta Singha,Aman Abhisek, Guru Kadam, Sabyasachi Ghosh and Hiranmaya Mishra (in progress).}
\def\cbmbook{B. Friman, C. Hohne, J. Knoll, S. Leupold, J. Randrup,R. Rapp. {\em etal}, {\em The CBM Physics Book: Compressed Baryonic
Matter in Laboratory Experiments. Lecture Notes in Physics, Springer, Berlin, Heidelberg,2011}.}
\def\jinrwhite{D. Blaschke, J. Aichelin, E. Bratkovskaya, V. Friese 
{\em etal}{\it Topical issues on exploring strongly interacting matter 
at high densities- nica white paper}, {\EPJA{52}{267}{2016}}.}
\def\rishi{Sreemoyee Sarkar and Rishi Sharma, arXiv: 17010001[hep-ph].}
\def\shenke{C. Gale, S. Jeon, B. Schenke, Int. J. Mod. Phys. {\bf A 28}, 134011,
(2013).}
\def\chamel{N. Chamel and P. Hansel, Living Rev. Rel. {\bf 11}, 10 (2008), arXiv:0812.3955[astro-ph]}
\def\sreddy{D. Page and S. Reddy, arXiv:1201.5602[nucl-th].}
\def\bjschaefer{B. J. Schaefer, J. M. Pawlowski and J. Wambach, {\PRD{76}{074023}{2007}}.}

\def\rischkeqm{O. Scavenius, A. Mocsy, I. N. Mishustin, D. H. Rischke,{\PRC{64}{045202}{2001}}.}
\def\guptatiwari{U.S. Gupta, V.K. Tiwari,{\PRD{85}{014010}{2012}}.}
\def\bielich{B.W. Mintz, R.Stiele, R.O. Ramos, J.S. Bielich,{\PRD{87}{036004}{2013}}}
\def\ranjita{H. Mishra, R.K. Mohapatra,{\PRD{95}{094014}{2017}}.}
\def\ghoshraha{S.K. Ghosh,A. Lahiri, S. Majumder, M.G. Mustafa, S. Raha, R. Ray,{\PRD{90}{054030}{2014}}}
\def\buballa{S. Carignano, M. Buballa, W.Elkamhawy,{\PRD{94}{034023}{2016}}}
\def\pisarskijhep{Yoshimasa Hidaka, Shu Lin, Robert D. Pisarski and Daisuke Satow, JHEP{\bf 10},005, (2015).}
\def\lata{Lata Thakur, P.K.Srivastava,Guru Prakash Kadam, Manu George, Hiranmaya Mishra,{\PRD{95}{096009}{2017}}}
\def\guru1{Guru Prakash Kadam, Hiranmaya Mishra,{\PRC{93}{025205}{2016}}}
\def\guru{Guru Prakash Kadam, Hiranmaya Mishra,{\PRC{92}{035203}{2015}};ibid,{\PRC{93}{025205}{2016}}; L. Thakur, P.K. Srivastava,
G. Kadam, M. George,and H. Mishra,{\PRD{95}{096009}{2017}}.}

\end{document}